\documentclass[submission,copyright,creativecommons]{eptcs}
\providecommand{\event}{COS 2013}

\providecommand{\urlalt}[2]{\href{#1}{#2}} 
\providecommand{\doi}[1]{doi:\urlalt{http://dx.doi.org/#1}{#1}} 

\usepackage{breakurl}            

\usepackage{enumerate}
\usepackage{amssymb}
\usepackage{amsthm}
\usepackage{hyperref}
\usepackage{mathpartir}
\usepackage{graphicx}
\usepackage{times}

\newtheorem{theorem}{Theorem}
\newtheorem{lemma}{Lemma}
\newtheorem{definition}{Definition}

\newcommand{\ie}{i.e., }
\newcommand{\eg}{e.g.}

\makeatletter
\def\verbatim{\small \@verbatim \frenchspacing\@vobeyspaces \@xverbatim}

\makeatother

\def\doframeit#1{\vbox{%
  \hrule height\fboxrule
    \hbox{%
      \vrule width\fboxrule \kern\fboxsep
      \vbox{\kern\fboxvsep #1\kern\fboxvsep }%
      \kern\fboxsep \vrule width\fboxrule }%
    \hrule height\fboxrule }}

\def\frameit{\smallskip \advance \linewidth by -7.5pt \setbox0=\vbox \bgroup
\strut \ignorespaces }

\def\endframeit{\ifhmode \par \nointerlineskip \fi \egroup
\doframeit{\box0}}

\newdimen \fboxvsep
\fboxvsep=\fboxsep
\advance \fboxsep by -3.5pt

\newcommand{\Mid}{\;\mid\;}

\newcommand{\lambdaf}{\ensuremath{\lambda^{F}_v}}
\newcommand{\lambdafdel}{\ensuremath{\lambda^{F}_{\rawshift,v}}}
\newcommand{\tm}{\ensuremath{t}}
\newcommand{\tmr}{\ensuremath{r}}
\newcommand{\tms}{\ensuremath{s}}

\newcommand{\tmzero}{\ensuremath{\tm_0}}
\newcommand{\tmone}{\ensuremath{\tm_1}}
\newcommand{\val}{\ensuremath{v}}
\newcommand{\valzero}{\ensuremath{\val_0}}
\newcommand{\valone}{\ensuremath{\val_1}}
\newcommand{\redex}{\ensuremath{r}}

\newcommand{\varx}{\ensuremath{x}}
\newcommand{\vark}{\ensuremath{k}}

\newcommand{\rawcallcc}{\mathcal{K}}
\newcommand{\callcc}[2]{\ensuremath{\rawcallcc{#1}.{#2}}}

\newcommand{\app}[2]{\ensuremath{#1\:#2}}

\newcommand{\throw}[2]{\ensuremath{{#1}\hookleftarrow{#2}}}

\newcommand{\throwctx}[2]{\ensuremath{{#1}\hookleftarrow{#2}}}
\newcommand{\throwctxp}[2]{\ensuremath{(\throwctx{#1}{#2})}}

\newcommand{\subst}[3]{\ensuremath{#1\{#3/#2\}}}

\newcommand{\substsimsim}[7]{\ensuremath{#1\{#3/#2,#5/#4,#7/#6\}}}

\newcommand{\reset}[1]{\langle{#1}\rangle}

\newcommand{\rawshift}{\ensuremath{\mathcal{S}}}
\newcommand{\shift}[2]{\ensuremath{\rawshift{#1}.{#2}}}

\newcommand{\lamtp}[3]{\ensuremath{\lambda{#1}^{#2}.#3}}
\newcommand{\lamtpp}[3]{\ensuremath{(\lambda{#1}^{#2}.#3)}}
\newcommand{\Lam}[2]{\ensuremath{\Lambda{#1}.{#2}}}
\newcommand{\Lamp}[2]{\ensuremath{(\Lambda{#1}.{#2})}}
\newcommand{\apptp}[2]{\ensuremath{#1\{#2\}}}

\newcommand{\ctx}{\ensuremath{\mathit{E}}}
\newcommand{\ctxzero}{\ensuremath{\mathit{E_0}}}
\newcommand{\ctxone}{\ensuremath{\mathit{E_1}}}
\newcommand{\hole}{\ensuremath{[\:]}}
\newcommand{\mtctx}{\ensuremath{\hole}}
\newcommand{\vctx}[2]{\ensuremath{#1\;#2}}
\newcommand{\vctxp}[2]{\ensuremath{(\vctx{#1}{#2})}}
\newcommand{\apctx}[2]{\ensuremath{#1\;#2}}

\newcommand{\inctx}[2]{\ensuremath{#1[#2]}}

\newcommand{\lamctx}[2]{\ensuremath{#1\;#2}}
\newcommand{\aptpctx}[2]{\ensuremath{#1\;\{#2\}}}

\newcommand{\reify}[1]{\ulcorner\! #1 \urcorner}

\newcommand{\metactx}{\mathit{F}}

\newcommand{\mtmetactx}{\ensuremath{\bullet}}
\newcommand{\ctxmetactx}[2]{\ensuremath{#1\:\#\: #2}}

\newcommand{\pgm}{\ensuremath{p}}

\newcommand{\program}[2]{\ensuremath{#2[#1]}}
\newcommand{\programDF}[3]{\ensuremath{#3[\langle{#2}[{#1}]\rangle]}}

\newcommand{\rawplug}{\ensuremath{\mathit{plug}}}
\newcommand{\plug}[2]{\ensuremath{\rawplug\;(#1,#2)}}
\newcommand{\rawplugmeta}{\ensuremath{\mathit{plug_m}}}
\newcommand{\plugmeta}[2]{\ensuremath{\rawplugmeta\;(#1,#2)}}

\newcommand{\isdef}{:=}

\newcommand{\rawftv}{\ensuremath{\mathit{ftv}}}
\newcommand{\ftv}[1]{\ensuremath{\rawftv\,(#1)}}

\newcommand{\redcbv}{\ensuremath{\rightarrow_{\rm v}}}

\newcommand{\rtclo}[1]{\ensuremath{#1^*}}
\newcommand{\relevalcbv}{\ensuremath{\rtclo{\redcbv}}}
\newcommand{\redcbn}{\ensuremath{\rightarrow_{\rm n}}}

\newcommand{\ctyp}{\ensuremath{C}}
\newcommand{\tpa}{\ensuremath{S}}
\newcommand{\tpb}{\ensuremath{T}}
\newcommand{\tpc}{\ensuremath{U}}
\newcommand{\tpd}{\ensuremath{V}}
\newcommand{\tpe}{\ensuremath{W}}
\newcommand{\tpf}{\ensuremath{X}}
\newcommand{\tpg}{\ensuremath{Y}}

\newcommand{\basetp}{\ensuremath{X}} 
\newcommand{\arrowtp}[2]{\ensuremath{#1\rightarrow#2}}
\newcommand{\arrowtpDF}[4]{{#1}\:_{#3}\rightarrow_{\:#4}{#2}}
\newcommand{\foralltp}[2]{\ensuremath{\forall#1.#2}}
\newcommand{\foralltpDF}[4]{\ensuremath{\forall#1.#2^{\:#3,#4}}}

\newcommand{\tpctx}[1]{\ensuremath{\neg{#1}}}

\newcommand{\basetpi}[1]{\ensuremath{\basetp_{#1}}}
\newcommand{\basetpc}{\ensuremath{c}}

\newcommand{\oftp}[2]{\ensuremath{#1\!:\!#2}}
\newcommand{\tpenv}{\ensuremath{\Gamma}}
\newcommand{\mttpenv}{\ensuremath{\cdot}}
\newcommand{\tpenvc}{\ensuremath{\Delta}}
\newcommand{\mttpenvc}{\ensuremath{\cdot}}
\newcommand{\tpenvext}[3]{\ensuremath{{#1},\, {#2} \, : \, {#3}}}

\newcommand{\typj}[3]{\ensuremath{#1\:\vdash\:#2\, : \, #3}}
\newcommand{\typjctrl}[4]{\ensuremath{#1;#2\:\vdash\:#3\, : \, #4}}
\newcommand{\typejDF}[6]{\ensuremath{#1;#2\:|\:#3\:\vdash\:#4\,:\,#5\:|\:#6}}
\newcommand{\typejcDF}[4]{\ensuremath{#1;#2\:\vdash\:#3\, : \, #4}}

\newcommand{\tpctxDF}[2]{#1\rhd#2}
\newcommand{\tpmetactxDF}[1]{\neg{#1}}

\newcommand{\tpvarcbn}[3]{#1^{#2,#3}}

\newcommand{\tpcont}{C}
\newcommand{\tpmcont}{D}

\newcommand{\rawrnorm}{\ensuremath{\mathcal{N}}}
\newcommand{\rnorm}[1]{\ensuremath{\rawrnorm(#1)}}
\newcommand{\rawQterm}[1]{\ensuremath{\mathcal{Q}_{#1}}}
\newcommand{\Qterm}[2]{\ensuremath{\rawQterm{#1}(#2)}}
\newcommand{\rawQtermDF}[3]{\ensuremath{\mathcal{Q}_{#1}^{#2,#3}}}
\newcommand{\QtermDF}[4]{\ensuremath{\rawQtermDF{#1}{#2}{#3}(#4)}}

\newcommand{\redc}[1]{\ensuremath{\mathcal{RC}(#1)}}
\newcommand{\rc}{\ensuremath{\mathcal{R}}}
\newcommand{\rcs}{\ensuremath{\mathcal{S}}}
\newcommand{\rct}{\ensuremath{\mathcal{T}}}
\newcommand{\rawrccontDF}[2]{\ensuremath{\mathcal{C}_{#1,#2}}}
\newcommand{\rccontDF}[3]{\ensuremath{\rawrccontDF{#1}{#2}(#3)}}
\newcommand{\rawrcmetacont}[1]{\ensuremath{\mathcal{M}_{#1}}}
\newcommand{\rcmetacont}[2]{\ensuremath{\rawrcmetacont{#1}(#2)}}
\newcommand{\paramrc}[3]{\ensuremath{RED_{#1}[#3/#2]}}
\newcommand{\paramrcsim}[5]{\ensuremath{RED_{#1}[#3/#2,#5/#4]}}

\newcommand{\rawrccont}[1]{\ensuremath{\mathcal{C}_{#1}}}
\newcommand{\rccont}[2]{\ensuremath{\rawrccont{#1}(#2)}}
\newcommand{\rawQtermrc}[1]{\ensuremath{\mathcal{Q}_{#1}}}
\newcommand{\Qtermrc}[2]{\ensuremath{\rawQterm{#1}(#2)}}

\newcommand{\Forall}[2]{\forall{#1}.\,{#2}}
\newcommand{\Exists}[2]{\exists{#1}.\,{#2}}

\newcommand{\impl}{\rightarrow}

\newcommand{\vectm}[1]{\ensuremath{\vec{#1}}}

\title{Proving termination of evaluation\\ for System F with control
  operators}
\author{
Ma{\l}gorzata Biernacka
\institute{Institute of Computer Science\\ University of Wroc{\l}aw} 
\email{mabi@cs.uni.wroc.pl}
\and
Dariusz Biernacki
\institute{Institute of Computer Science\\ University of Wroc{\l}aw} 
\email{dabi@cs.uni.wroc.pl}
\and
Sergue\"i Lenglet
\institute{LORIA \\ Universit{\'e} de Lorraine}
\email{serguei.lenglet@univ-lorraine.fr}
\and
Marek Materzok
\institute{Institute of Computer Science\\ University of Wroc{\l}aw} 
\email{marek.materzok@cs.uni.wroc.pl}
}
\def\titlerunning{Proving termination of evaluation for System F with control
  operators}
\def\authorrunning{M. Biernacka, D. Biernacki, S. Lenglet \& M. Materzok}

\begin{document}

\maketitle

\begin{abstract}
We present new proofs of termination of evaluation in reduction
semantics (i.e., a small-step operational semantics with explicit
representation of evaluation contexts) for System F with control
operators. We introduce a modified version of Girard's proof method
based on reducibility candidates, where the reducibility predicates
are defined on values and on evaluation contexts as prescribed by the
reduction semantics format. We address both abortive control operators
({\it callcc}) and delimited-control operators ({\it shift} and {\it
  reset}) for which we introduce novel polymorphic type systems, and
we consider both the call-by-value and call-by-name evaluation
strategies.
\end{abstract}

\section{Introduction}

Termination of reductions is one of the crucial properties of typed
$\lambda$-calculi. When considering a $\lambda$-calculus as a
deterministic programming language, one is usually interested in
termination of reductions according to a given evaluation strategy,
such as call by value or call by name, rather than in more general
normalization properties. A convenient format to specify such
strategies is reduction semantics, \ie a form of operational semantics
with explicit representation of evaluation (reduction)
contexts~\cite{Felleisen-Friedman:FDPC3}, where the evaluation
contexts represent continuations~\cite{Danvy:ICFP08}. Reduction
semantics is particularly convenient for expressing non-local control
effects and has been most successfully used to express the semantics
of control operators such as {\it
  callcc}~\cite{Felleisen-Friedman:FDPC3}, or {\it shift} and {\it
  reset}~\cite{Biernacka-al:LMCS05}.

For simply-typed languages with control operators, it is common to
prove termination of evaluation (and of normalization in general) by
translating, in a reduction-preserving way, terms in the source
language to a target language for which the normalization property has
been established
before~\cite{Griffin:POPL90,Schwichtenberg:Marktoberdorf95}. Such
indirect proofs in general can be cumbersome and, as argued by Ikeda
and Nakazawa~\cite{Ikeda-Nakazawa:IPL06}, they can be error-prone.

In a previous
work~\cite{Biernacka-Biernacki:MFPS09,Biernacka-Biernacki:PPDP09}, it
has been shown that a context-based variant of Tait's proof method
based on reducibility predicates~\cite{Tait:JSL67,Troelstra:73} allows
for direct and concise proofs of termination of evaluation in
reduction semantics for the simply-typed $\lambda$-calculus with
control operators, be they abortive or delimited.  Unlike
translation-based proofs, the context-based proof method directly
takes advantage of the format of the reduction semantics, where the
key role is played by evaluation contexts. So, for instance, in order
to prove termination of evaluation for the simply-typed
$\lambda$-calculus under call by value using the context-based method,
one defines mutually inductively reducibility predicates on values
(normal forms) as well as on evaluation contexts. The termination
result then follows by induction on well-typed terms, where the
reasoning is driven by the control flow of a typical evaluator in
continuation-passing
style~\cite{Biernacka-Biernacki:MFPS09,Biernacka-Biernacki:PPDP09}.

In this article, we show that the context-based method can be
generalized from reducibility predicates to reducibility candidates,
and therefore it provides simple proofs of termination of evaluation
for System F with control operators. Just as for the simply-typed
$\lambda$-calculi, normalization for polymorphic $\lambda$-calculi
with control operators has been mainly established indirectly, via
translations to strongly normalizing calculi: Harper and Lillibridge
reduced termination of call-by-value evaluation for $F_\omega$ with
{\it abort} and {\it callcc} to normalization in
$F_\omega$~\cite{Harper-Lillibridge:JFP96} by a CPS translation,
Parigot reduced strong normalization of the second-order
$\lambda\mu$-calculus to strong normalization of the simply-typed
$\lambda$-calculus~\cite{Parigot:JSL97}, Danos et al. reduced strong
normalization of the second-order classical logic to strong
normalization of linear logic~\cite{Danos-al:JSL97}, and Kameyama and
Asai reduced strong normalization of System F with {\it shift} and
{\it reset} under the standard semantics to strong normalization of
System F~\cite{Kameyama-Asai:SCSS08}.

On the other hand, Parigot directly proved strong normalization of the
second-order $\lambda\mu$-calculus using another variant of the
reducibility candidates~\cite{Parigot:JSL97}---we discuss this result
in Section~\ref{subsec:termination-abort}. Later on, further
adaptations of Tait-Girard's method have been proposed and applied to
various flavors of second-order
logic~\cite{Danos-Krivine:CSL00,Krivine:PS09,Lengrand-Miquel:APAL08,Mellies-Vouillon:LICS05}.
In particular, following Girard, the techniques of orthogonality have
been used as a framework in which the concepts of the original
reducibility method can be phrased. The use of orthogonality induces
the notion of context which is understood as a sequence of
terms---roughly corresponding to call-by-name contexts. Reducibility
candidates are then defined in terms of TT-closed sets. In contrast to
this approach, we consider concrete evaluation strategies and our
contexts come directly from the reduction semantics (either
call-by-name or call-by-value, and the contexts can be layered in the
delimited-control case), and in particular our reducibility candidates
only contain values and are not TT-closed. Another related work is the
proof of strong normalization for Moggi's computational calculus given
by Lindley and Stark~\cite{Lindley-Stark:TLCA05} who have introduced
the operation of TT-lifting in order to interpret computational types.
This operation seems to correspond to our definition of reducibility
for one layer of (reduction) contexts (see
Section~\ref{subsec:termination-abort}). However, in strong
normalization, the notion of context is to be understood as a means of
syntactically splitting a term rather than ``the remaining
computation.'' In particular, we do not analyze the reducibility of
terms forming reduction contexts.

The calculi we consider are System F with {\it callcc} under call by
value and call by name as well as System F with {\it shift} and {\it
  reset} under call by value and call by name, in each case with the
standard semantics, where, unlike in the ML-like semantics, evaluation
does not proceed under polymorphic
abstraction~\cite{Harper-Lillibridge:JFP96}. The type system for {\it
  callcc} is inspired by that of Harper and
Lillibridge~\cite{Harper-Lillibridge:JFP96}, whereas the type systems
for {\it shift} and {\it reset} (one for each evaluation strategy) are
new and they generalize Asai and Kameyama's type
system~\cite{Asai-Kameyama:APLAS07} in that they allow for polymorphic
abstractions over arbitrary expressions, not only pure ones. It is
worth noting that, as in the simply-typed
case~\cite{Biernacka-Biernacki:MFPS09,Biernacka-Biernacki:PPDP09}, the
context-based proofs we present in this article have the structure of
an evaluator in continuation-passing style.

We would like to stress that the semantics we consider in this article
are not instances of abstract machines working on explicit
decompositions of terms, where the process of decomposition is built
in the transitions of the system. Instead, we rely on the higher-level
reduction semantics approach where the operations of decomposition and
recomposition of terms are left implicit. Consequently, the type
systems we consider are in the form of natural deduction rather than
in sequent calculus~\cite{Curien-Herbelin:ICFP00}.

The rest of this article is organized as follows. In
Section~\ref{sec:abortive}, we present System F with abortive control
operators and we prove termination of evaluation under the
call-by-value and call-by-name evaluation strategies for this
system. We also relate this result to Parigot's
work~\cite{Parigot:JSL97}.  In Section~\ref{sec:delimited}, we present
System F with delimited-control operators and we prove termination of
evaluation under call by value and call by name. We also relate our
type systems to Asai and Kameyama's~\cite{Asai-Kameyama:APLAS07}. In
Section~\ref{sec:conclusion}, we conclude.

\section{System F with abortive control operators}
\label{sec:abortive}

In this section, we present a context-based proof of termination for
call-by-value evaluation in System F extended with the control
operator {\it callcc}. We use a variant of Girard's method of
reducibility candidates, where in particular we define reducibility
predicates for reduction contexts.

\subsection{Syntax and Semantics}
\label{subsec:abortive-syntax-cbv}

We consider the explicitly typed System F under the call-by-value
reduction strategy, which we extend with the binder version of the
{\it callcc} operator (denoted $\rawcallcc$) and with a construct to
apply captured continuations $\throw {}{}$, similar to the {\it throw}
construct of SML/NJ \cite{Harper-al:JFP93}. We call this language
$\lambdaf$.

\noindent 
The syntax of terms, term types, and call-by-value contexts of
$\lambdaf$ is defined as follows:
$$
\begin{array}{rrcl}
  \textrm{Terms:} &\tm &::=& \varx \Mid \lamtp \varx \tpa \tm \Mid \app \tm \tm \Mid \Lam
  \basetp \tm \Mid \apptp \tm \tpa \Mid 
\callcc \vark \tm \Mid \throw \vark \tm
  \Mid \throw {\reify \ctx} \tm \\[1mm]
  \textrm{Term types:} &\tpa& ::=& \basetp \Mid \arrowtp \tpa \tpa \Mid \foralltp \basetp \tpa\\[1mm]
  \textrm{CBV contexts:}&\ctx& ::=& \mtctx \Mid \lamctx{\lamtpp \varx \tpa \tm} \ctx \Mid \apctx \ctx
  \tm \Mid \aptpctx \ctx \tpa \Mid \throwctx {\reify \ctx} \ctx\\[1mm]
  \textrm{Values:} &\val &::=& \lamtp \varx \tpa \tm \Mid \Lam \basetp \tm
\end{array}
$$

\noindent
We let $\varx$ range over term variables, $\vark$ range over
continuation variables, and $\basetp$ range over type variables, and
we assume the three sets of variables are pairwise disjoint.  We use
capital letters starting from $\tpa$ to denote term types.  The term
$\Lam \basetp \tm$ quantifies over the type variable $\basetp$, while
the term $\apptp \tm \tpa$ instantiates such quantification with type
$\tpa$. The term $\callcc \vark \tm$ denotes the \emph{callcc}
operator that binds a captured context (representing a continuation)
to the variable $k$ and makes it available in its body $t$.  In turn,
constructs of the form $\throw \vark \tm$ and $\throw {\reify \ctx}
\tm$ denote the operation of \emph{throwing} the term $\tm$ to a
continuation variable $\vark$ and to the captured context $\reify
\ctx$, respectively. The use of $\reify\cdot$ indicates that a context
is reified as a term, as opposed to its role as the representation of
the ``rest of the program.''

Expressions of the form $\throw {\reify \ctx} \tm$ are not allowed in
source programs (because we do not let programmers handle contexts
explicitly), but they may occur during evaluation. In the sequel, it
will be useful to distinguish the subset of \emph{plain terms},
\ie\ terms without any subterm of the form $\throw {\reify \ctx} \tm$.

An abstraction $\lamtp \varx \tpa \tm$ (resp., $\Lam \basetp \tm$,
$\callcc \vark \tm$) binds $\varx$ (resp., $\basetp$, $\vark$) in
$\tm$, and a type $\foralltp \basetp \tpa$ binds $\basetp$ in
$\tpa$. We write $\ftv \tpa$ for the set of free type variables
occurring in type $\tpa$, defined in the usual way. The definitions of
free term variables, free type variables, and free continuation
variables of a term are also standard. A term is \emph{closed} if it
does not have any free variable of any kind. We identify terms and
types up to $\alpha$-conversion of their bound variables.

The syntax of reduction contexts encodes the reduction strategy,
here---call by value. The contexts can be seen as ``terms with a
hole'', and are represented inside-out. Informally, $\mtctx$ denotes
the empty context, $\lamctx{\lamtpp \varx \tpa \tm} \ctx$ represents
$\inctx \ctx {\app {\lamtpp \varx \tpa \tm} \hole}$ with the hole
indicated by $\hole$, $\apctx \ctx \tm$ represents $\inctx \ctx {\app
  \hole \tm}$, $\aptpctx \ctx \tpa$ represents $\inctx \ctx {\apptp
  \hole \tpa}$, and $\throwctx{\reify {\ctx_0}}{\ctx}$ represents
$\inctx \ctx {\throwctx{\reify {\ctx_0}} \hole}$. A reduction context
is closed if and only if all its components (terms, types, or contexts) are
closed.  We make the meaning of contexts precise by defining a
function $\rawplug$ which maps a term and a context to the term which
is obtained by putting the term in the hole of the context:
\vspace{-1mm}
\begin{eqnarray*}
  \plug \tm \mtctx &=& \tm \\
  \plug {\tm_0} {\lamctx{\lamtpp \varx \tpa \tm} \ctx} &=& \plug {\app {\lamtpp
      \varx \tpa \tm} \tm_0} \ctx\\
  \plug \tmzero {\apctx \ctx \tmone}  &=& \plug {\app \tmzero \tmone} \ctx\\
  \plug \tm {\aptpctx \ctx \tpa} &=& \plug {\apptp \tm \tpa} \ctx\\
  \plug \tm {\throwctx {\reify {\ctx_0}}{\ctx}} &=& \plug {\throw {\reify {\ctx_0}} \tm} \ctx
\end{eqnarray*}
We write $\inctx \ctx \tm$ for the result of plugging $\tm$ in the
context $\ctx$ (\ie\ the result of $\plug \tm \ctx$).

A program $\pgm$ is a closed plain term. When $\pgm = \inctx \ctx
{\tm}$, we say that $\pgm$ decomposes into term $\tm$ in the context
$\ctx$. In general, a program can be decomposed into a term in a
context in more than one way. For example, the program $\app {\lamtpp
  \varx \tpa \tmzero} \tmone$ can be represented by term $\tmone$ in
context $\lamctx{\lamtpp \varx \tpa \tmzero}{\mtctx}$, or by term
$\lamtpp \varx \tpa \tmzero $ in context $\apctx{\mtctx}{\tmone}$, or
by term $\app {\lamtpp \varx \tpa \tmzero} \tmone$ in context
$\mtctx$.

The one-step reduction relation in the call-by-value strategy is
defined on programs by the following rules:
\[
\begin{array}{rcll}
\program {\app{\lamtpp \varx \tpa \tm} \val} \ctx &\redcbv& \program {\subst
   \tm \varx \val} \ctx & (\beta_v) \\[2mm]
  \program {\apptp{\Lamp \basetp \tm} \tpa} \ctx &\redcbv& \program {\subst
    \tm \basetp \tpa} \ctx & (\beta_T) \\[2mm]
 \program {\callcc \vark \tm} \ctx &\redcbv& \program {\subst
    \tm \vark {\reify \ctx}} \ctx & (\textit{callcc})\\[2mm]
 \program {\throw {\reify \ctxzero} \val} \ctxone &\redcbv& \program \val \ctxzero & (\textit{throw}_v)
\end{array}
\]
where $\subst \tm \varx \val$ (resp., $\subst \tm \basetp \tpa$,
$\subst \tm \vark {\reify \ctx}$) is the usual capture-avoiding
substitution of value $\val$ (resp., of type $\tpa$, of context
$\reify \ctx$) for variable $\varx$ (resp., for $\basetp$, for
$\vark$) in $\tm$. The rules $(\beta_v)$ and $(\beta_T)$ are standard
in System F; in addition, we introduce the rule $(\textit{callcc})$,
where the current context $\ctx$ is captured by the {\it callcc}
operator and bound to the continuation variable $\vark$, and the rule
$(\textit{throw}_v)$, where a previously captured context $\ctxzero$
is restored as the current context, and the context $\ctxone$ is
discarded (the latter fact shows the abortive character of the
\emph{callcc} operator). The plugged terms on the left-hand side of
the arrow in the above rules are called redexes, and are ranged over
by $\redex$. Note that the reduction relation is not compatible, \ie{}
it only applies to entire programs (due to the context capture in the
rule $(\textit{callcc})$).

We define the call-by-value evaluation relation as the reflexive and
transitive closure of the relation $\redcbv$. The expected result of
evaluation is a value.

The reduction relation $\redcbv$ is deterministic; this property is
ensured by the unique-decomposition lemma. We could state this lemma
in a general version for all terms, but in order to consider only
well-behaved programs and simplify the statement of the lemma, we
choose to postpone it to the next section where we define well-typed
programs.

\subsection{Type System} 

We define a type system for $\lambdaf$ that is an extension of the
type system for the lambda calculus introduced by Biernacka and
Biernacki~\cite{Biernacka-Biernacki:MFPS09}, where types are assigned
to terms as well as to contexts. The syntax of context types is
$\tpctx{\tpa}$. Roughly, the type $\tpctx \tpa$ of a context $\ctx$
indicates that any well-typed term of type $\tpa$ can be plugged in
$\ctx$. The answer type of a context need not be specified, and it is
often taken to be $\bot$ to reflect the fact that continuations never
return.\footnote{This decision has more serious implications when a
  type system is studied from a logical perspective via the
  Curry-Howard isomorphism (see for
  example~\cite{Ariola-al:HOSC07}). However, we do not take this
  viewpoint in this article.} The answer type of a closed evaluation
context $\ctx$ of type $\tpctx \tpa$ can be determined by typing the
expression $\inctx{\ctx}{x}$ for a fresh variable $x$ of type $\tpa$.

We let $\tpenv$ range over type environments for term variables
(\ie\ lists of pairs of the form $\oftp \varx \tpa$), and we let
$\tpenvc$ range over type environments for continuation variables
(\ie\ lists of pairs of the form $\oftp \vark \ctyp$). For $\tpenv =
\tpenvext {\oftp {\varx_1}{\tpa_1}, \ldots}{\varx_ n}{\tpa_n}$ and
$\tpenvc = \tpenvext {\oftp {\vark_1}{\tpctx{\tpa_1}}, \ldots}{\vark_
  n}{\tpctx{\tpa_n}}$, we define $\ftv \tpenv \isdef \cup_{i \in \{ 1,
  \ldots, n\} } \ftv{\tpa_i}$ and $\ftv \tpenvc \isdef \cup_{i \in \{
  1, \ldots, n\} } \ftv{\tpa_i}$. The typing rules for terms and
contexts are shown in Figure~\ref{fig:typing-cbv-callcc}.

\begin{figure*}[t!!]
\begin{flushleft}
Typing terms ($\tpa ::= \basetp \Mid \arrowtp \tpa \tpb \Mid \foralltp \basetp \tpa$):
\end{flushleft}
\begin{mathpar}
\inferrule{ }
          {\typjctrl{\tpenvext \tpenv \varx \tpa} \tpenvc \varx \tpa}
\and
\inferrule{\typjctrl{\tpenvext \tpenv \varx \tpa} \tpenvc \tm \tpb }
          {\typjctrl \tpenv \tpenvc {\lamtp \varx \tpa \tm}{\arrowtp \tpa \tpb}}
\and
\inferrule{\typjctrl \tpenv \tpenvc \tmzero {\arrowtp \tpa \tpb} \\ 
  \typjctrl \tpenv \tpenvc \tmone \tpa}
{\typjctrl \tpenv \tpenvc {\app \tmzero \tmone} \tpb}
\and
\inferrule{\typjctrl \tpenv \tpenvc \tm \tpa \\ \basetp \notin \ftv \tpenv \cup
  \ftv\tpenvc}
{\typjctrl \tpenv \tpenvc {\Lam \basetp \tm}{\foralltp \basetp \tpa}}
\and
\inferrule{\typjctrl \tpenv \tpenvc \tm {\foralltp \basetp \tpa}}
{\typjctrl \tpenv \tpenvc {\apptp \tm \tpb}{\subst \tpa \basetp \tpb}}
\and
\inferrule{\typjctrl \tpenv {\tpenvext \tpenvc \vark {\tpctx \tpa}} \tm \tpa}
{\typjctrl \tpenv \tpenvc {\callcc \vark \tm} \tpa} 
\and
\inferrule{\typjctrl \tpenv {\tpenvext \tpenvc \vark {\tpctx \tpa}} \tm \tpa}
{\typjctrl \tpenv {\tpenvext \tpenvc \vark {\tpctx \tpa}}{\throw \vark \tm} \tpb} 
\end{mathpar}
\begin{flushleft}
Typing contexts ($\tpcont ::= \tpctx \tpa$):
\end{flushleft}
\begin{mathpar}
  \inferrule{ }
  {\typjctrl \tpenv \tpenvc \mtctx {\tpctx \tpa}}
  \and
  \inferrule{\typjctrl \tpenv \tpenvc {\lamtp \varx \tpa \tm}{\arrowtp \tpa \tpb} \\ \typjctrl
    \tpenv \tpenvc \ctx {\tpctx \tpb}}
  {\typjctrl \tpenv \tpenvc {\lamctx{\lamtpp \varx \tpa \tm} \ctx} {\tpctx \tpa}}
  \and
  \inferrule{\typjctrl \tpenv \tpenvc \tm \tpa \\ \typjctrl \tpenv \tpenvc \ctx {\tpctx \tpb}}
  {\typjctrl \tpenv \tpenvc {\apctx \ctx \tm}{\tpctx {(\arrowtp \tpa \tpb)}} }
  \and
  \inferrule{\typjctrl \tpenv \tpenvc \ctx {\tpctx {(\subst \tpa \basetp \tpb)}}}
  {\typjctrl \tpenv \tpenvc {\aptpctx \ctx \tpb} {\tpctx {(\foralltp \basetp
        \tpa)}}}
\and
\inferrule{\typjctrl \tpenv \tpenvc \ctxzero {\tpctx \tpa}  \\ \typjctrl \tpenv
  \tpenvc \ctxone {\tpctx \tpb}}
{\typjctrl \tpenv \tpenvc {\throw {\reify \ctxzero} \ctxone} {\tpctx \tpa}} 
\end{mathpar}
\caption{Typing rules for System F with abortive control operators}
\label{fig:typing-cbv-callcc}
\end{figure*}

We can now state the unique-decomposition lemma that ensures the
determinism of the reduction relation $\redcbv$, and progress of
reduction:
\begin{lemma}[Unique decomposition]
  For all well-typed programs $\pgm$, $\pgm$ either is a value, or it
  decomposes uniquely into a context $\ctx$ and a redex $\redex$,
  \ie\ $\pgm = \program \redex \ctx$.
\end{lemma}

As for the subject reduction property, it is a more subtle issue. In
the reduction rule $(\textit{throw}_v)$, the current evaluation
context $\ctx_1$ is replaced by another context $\ctx_0$, where the
answer types of the two contexts do not have to be related in any
way. Therefore, as observed
before~\cite{Biernacka-Biernacki:MFPS09,Harper-al:JFP93,Wright-Felleisen:IaC94},
in general subject reduction does not hold for languages with
reduction and typing rules for continuation invocation similar to the
ones presented in this work. However, if we assume that all reified
contexts in a given term have the same answer type as the type of the
term itself---which is the case for all terms in the reduction
sequence starting in a plain term---subject reduction will be
recovered~\cite{Biernacka-Biernacki:MFPS09,Harper-al:JFP93,Wright-Felleisen:IaC94}. Such
an assumption can be made implicit~\cite{Harper-al:JFP93} or explicit
in a refined type system that controls the answer types of evaluation
contexts~\cite{Biernacka-Biernacki:MFPS09,Wright-Felleisen:IaC94}. It
can be shown that from subject reduction of such a refined type system
strong type soundness for plain terms in the original type system
follows~\cite{Biernacka-Biernacki:MFPS09,Wright-Felleisen:IaC94}.

\subsection{Termination}
\label{subsec:termination-abort}

We now prove termination of the call-by-value evaluation for
$\lambdaf$, using a context variant of Girard's method of reducibility
candidates~\cite{Girard-al:89}. Our definition of a reducibility
candidate is simpler than in Girard's proof of strong normalization
for system F, because we are interested only in the termination of the
call-by-value evaluation, not in strong normalization. Moreover, we
exploit the structure of the call-by-value continuation-passing
style~\cite{Griffin:POPL90,Harper-Lillibridge:JFP96,Plotkin:TCS75}
that underlies the semantics of the language we consider, and
therefore the central role in the proof is played by predicates on
evaluation contexts (representing continuations) and on values, and
not on arbitrary terms.

First, we define the normalization predicate $\rnorm \pgm$ as follows:
\[\rnorm \pgm \isdef \Exists \val {\pgm \relevalcbv \val},\] \ie\ a program $\pgm$ normalizes ($\rnorm\pgm$ holds) if
it reduces in several steps to a value.
\begin{definition}[Reducibility candidate]
A \emph{reducibility candidate} $\rc$ of type $\tpa$ is any set of
closed values of type $\tpa$.
\end{definition}
We write $\redc \tpa$ for the set of reducibility candidates of type
$\tpa$. For each reducibility candidate $\rc$ of type $\tpa$, we
define the associated predicate $\rawrccont \rc$ on closed contexts of
type $\tpctx \tpa$ as follows:
\[
\rccont \rc \ctx \isdef \Forall \val {\val \in \rc \impl \rnorm {\program \val
    \ctx}}
\]

As in the original proof, we introduce the notion of parametric
reducibility candidates. However, we base our definition on the CPS
interpretation of terms rather than on the direct-style
interpretation. Let $\tpa$ be a type, $\ftv \tpa \subseteq \vectm
\basetp$
\footnote{Henceforth, for any metavariable $m$, we write $\vectm m$ to range over
 sequences of entities denoted by $m$.}
, $\vectm \tpb$ be a sequence of types of the same size as $\vectm
\basetp$, and $\vectm \rc$ be such that $\rc_i \in \redc{\tpb_i}$. We
define the parametric reducibility candidate $\paramrc \tpa {\vectm
  \basetp}{\vectm \rc}$ by induction on $\tpa$:
\[
\begin{array}{rcl}
  \val \in \paramrc{\basetpi i}{\vectm \basetp}{\vectm \rc} & \mbox{ iff } & \val \in \rc_i 
  \\
  \valzero \in \paramrc {\arrowtp {\tpa_1}{\tpa_2}}{\vectm \basetp}{\vectm \rc}
  & \mbox{ iff } 
  & \forall \valone . \valone \in \paramrc {\tpa_1}{\vectm \basetp}{\vectm \rc}
    \impl 
    \Forall \ctx {\rccont {\paramrc {\tpa_2}{\vectm \basetp}{\vectm
          \rc}} \ctx \impl \rnorm {\program {\app \valzero \valone} \ctx}} 
  \\
  \val \in \paramrc {\foralltp \basetpc \tpa}{\vectm \basetp}{\vectm \rc} &\mbox
       { iff }& 
       \forall \tpc. 
       \forall \rcs. \rcs \in \redc \tpc \impl 
       \Forall \ctx {\rccont
             {\paramrcsim \tpa {\vectm \basetp}{\vectm \rc} \basetpc \rcs} \ctx \impl
             \rnorm {\program {\apptp \val \tpc} \ctx}}
\end{array}
\]
It is easy to see that $\paramrc \tpa {\vectm \basetp}{\vectm \rc} \in
\redc{\subst \tpa {\vectm \basetp}{\vectm \tpb}}$. To prove the main
result, we need a substitution lemma.

\begin{lemma}
  \label{l:paramrc-subst}
  We have $\paramrc {\subst \tpa \basetpc \tpb}{\vectm \basetp}{\vectm
    \rc}=\paramrcsim \tpa{\vectm \basetp}{\vectm \rc} \basetpc {\paramrc \tpb {\vectm \basetp}{\vectm \rc}}$.
\end{lemma}

\proof
  By induction on $\tpa$.
\qed

We are now ready to state the main lemma:

\begin{lemma}
  \label{l:termination-cbv-abort}
  Let $\tm$ be a plain term such that $\typjctrl \tpenv \tpenvc \tm
  \tpa$, $\tpenv = \tpenvext {\oftp {\varx_1}{\tpb_1},\:
    \ldots}{\varx_n}{\tpb_n}$, and $\tpenvc = \tpenvext {\oftp
    {\vark_1}{\tpctx {\tpc_1}},\: \ldots}{\vark_m}{\tpctx
    {\tpc_m}}$. Let $\{ \basetpi 1, \ldots, \basetpi p \} = \ftv \tpa
  \cup \ftv \tpenv \cup \ftv \tpenvc$.  Let $\vectm \tpd$ be a
  sequence of types of length $p$, and $\vectm \rc$ be reducibility
  candidates such that $\rc_i \in \redc{\tpd_i}$ for all
  $i=1,\ldots,p$. Let $\vectm \val$ be closed values such that $\typj
  {\mttpenv;\mttpenvc} {\val_i}{\subst {\tpb_i}{\vectm \basetp}{\vectm
      \tpd}}$ and $\val_i \in \paramrc{\tpb_i}{\vectm \basetp}{\vectm
    \rc}$ for all $i=1,\ldots,n$. Let $\vectm{\ctx}$ be closed
  contexts such that
  $\typj{\mttpenv;\mttpenvc}{\ctx_i}{\tpctx{\tpc_i}}$ and $\rccont
  {\paramrc {\tpc_i}{\vectm \basetp}{\vectm \rc}}{\ctx_i}$ for all
  $i=1,\ldots,m$, and let $\ctx$ be such that
  $\typj{\mttpenv;\mttpenvc}{\ctx}{\tpctx{\tpa}}$ and $\rccont
  {\paramrc \tpa {\vectm \basetp}{\vectm \rc}} \ctx$. Then $\rnorm
  {\program {\substsimsim \tm {\vectm \basetp}{\vectm \tpd}{\vectm
        \varx}{\vectm \val}{\vectm \vark}{\vectm {\reify \ctx}}}
    \ctx}$ holds.
\end{lemma}

\proof
  By induction on $\tm$.
\begin{itemize}
\item In the case $\tm = \varx_i$, we have $\substsimsim \tm {\vectm
  \basetp}{\vectm \tpd}{\vectm \varx}{\vectm \val}{\vectm
  \vark}{\vectm {\reify \ctx}}=\val_i$, as well as
  $\tpb_i=\tpa$. Because $\val_i \in \paramrc{\tpb_i}{\vectm
  \basetp}{\vectm \rc}$, by definition of $\rccont
  {\paramrc{\tpb_i}{\vectm \basetp}{\vectm \rc}} \ctx$, we have the
  required result. 
\item
  In the case $\tm = \lamtp{\varx}{\tpa_1}{\tms}$, we have $\tpa =
  \arrowtp{\tpa_1}{\tpa_2}$. Let $\tms'=\substsimsim \tms {\vectm \basetp}{\vectm
    \tpd}{\vectm \varx}{\vectm \val}{\vectm \vark}{\vectm {\reify \ctx}}$ and $\tpa'_1 =
  \subst {\tpa_1}{\vectm \basetp}{\vectm \tpd}$; then $\substsimsim \tm {\vectm
    \basetp}{\vectm \tpd}{\vectm \varx}{\vectm \val}{\vectm \vark}{\vectm {\reify \ctx}}
  = \lamtp \varx {\tpa'_1}{\tms'}$. We now prove that $\lamtp \varx
  {\tpa'_1}{\tms'} \in \paramrc \tpa {\vectm \basetp}{\vectm \rc}$; from
  that we can deduce the required result by the definition of $\rccont {\paramrc
    \tpa {\vectm \basetp}{\vectm \rc}} \ctx$. Let $\val$ be such that $\val
  \in \paramrc {\tpa_1}{\vectm \basetp}{\vectm \rc}$, and let $\ctx'$ be such that
  $\rccont {\paramrc {\tpa_2}{\vectm \basetp}{\vectm \rc}}{\ctx'}$. We have
  $\program {\app{\lamtpp \varx {\tpa'_1}{\tms'}} \val}{\ctx'} \redcbv \program
  {\subst {\tms'} \varx \val}{\ctx'}$. By the induction hypothesis, we have $\rnorm{\program
    {\subst {\tms'} \varx \val}{\ctx'}}$, therefore 
  $\rnorm{\program{\app{\lamtpp \varx {\tpa'_1}{\tms'}}
      \val}{\ctx'}}$ holds. Consequently, $\lamtp \varx {\tpa'_1}{\tms'}
  \in \paramrc \tpa {\vectm \basetp}{\vectm \rc}$ as required.
\item
  In the case $\tm = \app \tmzero \tmone$, we have $\typjctrl \tpenv
  \tpenvc \tmzero {\arrowtp {\tpa'} \tpa}$ and $\typjctrl \tpenv
  \tpenvc \tmone {\tpa'}$ for some $\tpa'$. Let $\tmzero' =
  \substsimsim \tmzero {\vectm \basetp}{\vectm \tpd}{\vectm
    \varx}{\vectm \val}{\vectm \vark}{\vectm {\reify \ctx}}$, $\tmone'
  = \substsimsim \tmone {\vectm \basetp}{\vectm \tpd}{\vectm
    \varx}{\vectm \val}{\vectm \vark}{\vectm {\reify \ctx}}$; then
  $\substsimsim \tm {\vectm \basetp}{\vectm \tpd}{\vectm \varx}{\vectm
    \val}{\vectm \vark}{\vectm {\reify \ctx}} = \app
  {\tmzero'}{\tmone'}$. We then have $\program {\app {\tmzero'}{\tmone'}}
  \ctx = \program {\tmzero'}{\apctx \ctx {\tmone'}}$, and to conclude,
  we would like to apply the induction hypothesis to $\tmzero$. To
  this end, we have to prove that $\rccont{\paramrc{\arrowtp {\tpa'}
      \tpa}{\vectm \basetp}{\vectm \rc}}{\apctx \ctx {\tmone'}}$
  holds. Let $\valzero$ be such that $\valzero \in \paramrc{\arrowtp
    {\tpa'} \tpa}{\vectm \basetp}{\vectm \rc}$. We want to prove that
  $\rnorm {\program \valzero {\apctx \ctx {\tmone'}}}$ holds, which is
  equivalent to proving that $\rnorm {\program {\tmone'}{\vctxp
      \valzero \ctx}}$ holds. Again, we want to prove this fact by
  using the induction hypothesis on $\tmone$, but to do this, we first
  have to prove that $\rccont{\paramrc{\tpa'}{\vectm \basetp}{\vectm
      \rc}}{\vctx \valzero \ctx}$ holds. Let $\valone$ be such that
  $\valone \in \paramrc{\tpa'}{\vectm \basetp}{\vectm \rc}$. Since we
  have $\valzero \in \paramrc{\arrowtp {\tpa'} \tpa}{\vectm
    \basetp}{\vectm \rc}$ and $\rccont{\paramrc{\tpa}{\vectm
      \basetp}{\vectm \rc}} \ctx$, therefore $\rnorm {\program {\app
      \valzero \valone} \ctx}$ holds, \ie we have $\rnorm {\program
    \valone {\vctxp \valzero \ctx}}$. Consequently,
  $\rccont{\paramrc{\tpa'}{\vectm \basetp}{\vectm \rc}}{\vctx \valzero
    \ctx}$ holds. Therefore, we can use the induction hypothesis on
  $\tmone$ to deduce that $\rnorm {\program {\tmone'}{\vctxp \valzero
      \ctx}}$ holds. As a result, we therefore have
  $\rccont{\paramrc{\arrowtp {\tpa'} \tpa}{\vectm \basetp}{\vectm
      \rc}}{\apctx \ctx {\tmone'}}$ and we can prove the required fact
  by using the induction hypothesis on $\tmzero$.
\item
  In the case $\tm = \Lam \basetpc \tms$, $\tpa = \foralltp \basetpc
  {\tpa'}$ for some $\tpa'$. Let $\tms'= \substsimsim \tms {\vectm
    \basetp}{\vectm \tpd}{\vectm \varx}{\vectm \val}{\vectm
    \vark}{\vectm {\reify \ctx}}$ and $\substsimsim \tm {\vectm
    \basetp}{\vectm \tpd}{\vectm \varx}{\vectm \val}{\vectm
    \vark}{\vectm {\reify \ctx}} = \Lam \basetpc {\tms'}$. We now
  prove that $\Lam \basetpc {\tms'} \in \paramrc \tpa {\vectm
    \basetp}{\vectm \rc}$; the required result then holds by the
  definition of $\rccont {\paramrc \tpa {\vectm \basetp}{\vectm \rc}}
  \ctx$. Let $\tpd'$ be a type and let $\rc' \in \redc {\tpd'}$. Let
  $\ctx'$ be such that $\rccont {\paramrcsim {\tpa'}{\vectm
      \basetp}{\vectm \rc} \basetpc {\rc'}}{\ctx'}$ holds. We have
  $\program {\apptp {\Lamp \basetpc {\tms'}}{\tpd'}}{\ctx'} \redcbv
  \program {\subst {\tms'} \basetpc {\tpd'}}{\ctx'}$. By the induction
  hypothesis, we have that $\rnorm {\program {\subst {\tms'} \basetpc
      {\tpd'}}{\ctx'}}$ holds, therefore we obtain that $\rnorm
  {\program {\apptp {\Lamp \basetpc {\tms'}}{\tpd'}}{\ctx'}}$ holds as
  required.
\item
  In the case $\tm = \apptp{\tmzero}{\tpd'}$, we have $\typjctrl
  \tpenv \tpenvc \tmzero {\foralltp \basetpc {\tpa'}}$ with $\tpa =
  \subst {\tpa'} \basetpc {\tpd'}$ for some $\tpa'$. Let $\tmzero' =
  \substsimsim \tmzero {\vectm \basetp}{\vectm \tpd}{\vectm
    \varx}{\vectm \val}{\vectm \vark}{\vectm {\reify \ctx}}$ and
  $\tpd'' = \subst {\tpd'}{\vectm \basetp}{\vectm \tpd}$; we have
  $\substsimsim \tm {\vectm \basetp}{\vectm \tpd}{\vectm \varx}{\vectm
    \val}{\vectm \vark}{\vectm {\reify \ctx}} = \apptp
               {\tmzero'}{\tpd''}$, $\program {\apptp
                 {\tmzero'}{\tpd''}} \ctx = \program
               {\tmzero'}{\aptpctx \ctx {\tpd''}}$. To conclude, we
               want to apply the induction hypothesis to $\tmzero$,
               but first we have to prove that
               $\rccont{\paramrc{\foralltp \basetpc {\tpa'}}{\vectm
                   \basetp}{\vectm \rc}}{\aptpctx \ctx {\tpd''}}$
               holds. Let $\val$ be such that $\val \in
               \paramrc{\foralltp \basetpc {\tpa'}}{\vectm
                 \basetp}{\vectm \rc}$. By
               Lemma~\ref{l:paramrc-subst}, $\paramrc {\subst {\tpa'}
                 \basetpc {\tpd'}}{\vectm \basetp}{\vectm
                 \rc}=\paramrcsim {\tpa'}{\vectm \basetp}{\vectm \rc}
               \basetpc {\paramrc {\tpd'} {\vectm \basetp}{\vectm
                   \rc}}$ and hence $\rccont {\paramrcsim
                 {\tpa'}{\vectm \basetp}{\vectm \rc} \basetpc
                 {\paramrc {\tpd'}{\vectm \basetp}{\vectm \rc}}} \ctx$
               holds. Besides, we have that $\paramrc {\tpd'} {\vectm
                 \basetp}{\vectm \rc} \in \redc{\tpd''}$ holds, so by
               the definition of $\paramrc{\foralltp \basetpc
                 {\tpa'}}{\vectm \basetp}{\vectm \rc}$ we obtain
               $\rnorm {\program {\apptp \val {\tpd''}} \ctx}$, \ie
               $\rnorm {\program \val {\aptpctx \ctx {\tpd''}}}$
               holds. Therefore, $\rccont{\paramrc{\foralltp \basetpc
                   {\tpa'}}{\vectm \basetp}{\vectm \rc}}{\aptpctx \ctx
                 {\tpd''}}$ holds, hence we have the required result
               by using the induction hypothesis on $\tms$.
\item
  Suppose $\tm = \callcc \vark \tms$. Let $\tms'= \substsimsim \tms
  {\vectm \basetp}{\vectm \tpd}{\vectm \varx}{\vectm \val}{\vectm
    \vark}{\vectm {\reify \ctx}}$. We then have the equality
  $\substsimsim \tm {\vectm \basetp}{\vectm \tpd}{\vectm \varx}{\vectm
    \val}{\vectm \vark}{\vectm {\reify \ctx}} = \callcc \vark
  {\tms'}$. Since $\program {\callcc \vark {\tms'}} \ctx \redcbv
  \program {\subst {\tms'} \vark {\reify \ctx}} \ctx$, we obtain
  $\rnorm {\program {\subst {\tms'} \vark {\reify \ctx}} \ctx}$ by the
  induction hypothesis, and therefore $\rnorm {\program {\callcc \vark
      {\tms'}} \ctx}$ holds.
\item  
  Suppose $\tm = \throw {\vark_i} \tms$. Let $\tms'= \substsimsim \tms
  {\vectm \basetp}{\vectm \tpd}{\vectm \varx}{\vectm \val}{\vectm
    \vark}{\vectm {\reify \ctx}}$. We then have the equality
  $\substsimsim \tm {\vectm \basetp}{\vectm \tpd}{\vectm \varx}{\vectm
    \val}{\vectm \vark}{\vectm {\reify \ctx}} = \throw
  {\reify{\ctx_i}}{\tms'}$. The program $\program {\throw {\reify
      {\ctx_i}}{\tms'}} \ctx$ is equivalent to $\program
  {\tms'}{\throwctxp {\reify{\ctx_i}} \ctx}$, so we need to prove that $\rnorm
  {\program {\tms'}{\throwctxp {\reify{\ctx_i}} \ctx}}$ holds, applying
  the induction hypothesis to $\tms$. To this end, we first prove that
  $\rccont {\paramrc{\tpc_i}{\vectm \basetp}{\vectm \rc}}{\throwctx
    {\reify{\ctx_i}} \ctx}$ holds. Let $\val$ be such that $\val \in
  \paramrc{\tpc_i}{\vectm \basetp}{\vectm \rc}$. The program $\program
  \val {\throwctxp {\reify{\ctx_i}} \ctx}$ is equivalent to $\program
       {\throw {\reify{\ctx_i}} \val} \ctx$. We have $\program {\throw
         {\reify{\ctx_i}} \val} \ctx \redcbv \program \val {\ctx_i}$,
       and since $\rccont {\paramrc{\tpc_i}{\vectm \basetp}{\vectm
           \rc}}{\ctx_i}$ holds, we obtain $\rnorm {\program \val
         {\ctx_i}}$. Consequently, $\rnorm {\program \val {\throwctxp
           {\reify{\ctx_i}} \ctx}}$ holds.
\end{itemize}
\qed

\begin{theorem}
\label{thm:termination-cbv-callcc}
  If $\pgm$ is a well-typed program, then $\rnorm{\pgm}$ holds.
\end{theorem}

\proof We have $\rccont \rc \mtctx$ for any $\rc$, therefore we can
use the previous lemma.  \qed

The proof of Theorem~\ref{thm:termination-cbv-callcc} is constructive
and its computational content is a call-by-value evaluator for plain
terms in the continuation-passing style that is an instance of
normalization by
evaluation~\cite{Biernacka-Biernacki:MFPS09,Biernacka-Biernacki:PPDP09}.

\subsection{Call by name}
\label{subsec:callbyname-abort}

The proof method can be adapted to the call-by-name strategy, again by
using a corresponding con\-ti\-nuation-passing style interpretation of
terms~\cite{Griffin:POPL90,Harper-Lillibridge:JFP96,Plotkin:TCS75}. In
this case, the syntax of reduction contexts becomes:
$$
\begin{array}{llll}
  \textrm{CBN contexts:}&\ctx& ::=& \mtctx \Mid \apctx \ctx \tm \Mid
  \aptpctx \ctx \tpa
\end{array}
$$ and the reduction rules are modified in that a lambda abstraction
and a throwing operation can be applied to an arbitrary term instead
of only to a value, in the rules $(\beta_n)$ and $(\textit{throw}_n)$
below:
\[
\begin{array}{rcll}
  \program {\app{\lamtpp \varx \tpa \tmzero} \tmone} \ctx &\redcbn& \program {\subst
    \tmzero \varx \tmone} \ctx & (\beta_n)\\[2mm]
  \program {\apptp{\Lamp \basetp \tm} \tpa} \ctx &\redcbn& \program {\subst
    \tm \basetp \tpa} \ctx & (\beta_T)\\[2mm]
  \program {\callcc \vark \tm} \ctx &\redcbn& \program {\subst
    \tm \vark {\reify \ctx}} \ctx & (\textit{callcc})\\[2mm]
  \program {\throw {\reify \ctxzero} \tm} \ctxone &\redcbn& \program \tm \ctxzero & (\textit{throw}_n)
\end{array}
\]

The type system is as before, except there are fewer rules for typing
contexts.  In the remainder of this section we briefly point out the
main differences in the proof of termination of evaluation between the
call-by-value and the call-by-name strategies.

A reducibility candidate of type $\tpa$ in call by name is a set of
values of type $\tpa$ (where values are as in call by value), and the
definition of the associated predicate $\rawrccont \rc$ is the same as
in call by value, except that the predicate $\rnorm\cdot$ is defined
using the call-by-name reduction relation $\redcbn$.

However, parametric reducibility candidates are defined in a
substantially different way, reflecting the call-by-name strategy.
Let $\tpa$ be a type, $\ftv\tpa \subseteq \vectm \basetp$, $\vectm
\tpb$ be a sequence of types of the same size as $\vectm \basetp$, and
$\vectm \rc$ be reducibility candidates such that each $\rc_i$ is of
type $\tpb_i$. We define the parametric reducibility candidate
$\paramrc \tpa {\vectm \basetp}{\vectm \rc}$ of type $\subst \tpa
{\vectm \basetp}{\vectm \tpb}$ as follows:
\[
\begin{array}{rcl}
  \val \in \paramrc {\basetpi i}{\vectm \basetp}{\vectm \rc} & \mbox{ iff } & \val \in
  \rc_i \\
  \val \in \paramrc {\arrowtp {\tpa_1}{\tpa_2}}{\vectm \basetp}{\vectm \rc}
  & \mbox{ iff }& \Forall \tm {\Qtermrc {\paramrc {\tpa_1}{\vectm
        \basetp}{\vectm \rc}} \tm \impl \Qtermrc {\paramrc {\tpa_2}{\vectm \basetp}{\vectm
          \rc}}{\app \val \tm}} \\
  \val \in \paramrc {\foralltp \basetpc \tpa}{\vectm \basetp}{\vectm \rc} &\mbox
  { iff }& \forall
  \tpc . \forall \rcs . \rcs \in \redc \tpc \impl 
  \Forall \ctx {\rccont
        {\paramrcsim \tpa {\vectm \basetp}{\vectm \rc} \basetpc \rcs} \ctx \impl \rnorm {\program {\apptp \val \tpc} \ctx}}
\end{array}
\]
with
\[
\Qtermrc \rc \tm  = \Forall \ctx {\rccont \rc \ctx \impl \rnorm {\program
      \tm \ctx}}
\]

\noindent
The predicate $\rawQtermrc{}$ used in the second clause of this
definition reflects the fact that a term given as argument to a
function is not yet a value (whose reducibility is immediate), but it
can be seen as a delayed computation that may be forced later, by
putting it in a context.

The main lemma is now formulated as follows:
\begin{lemma}
\label{lem:main-cbn-callcc}
  Let $\tm$ be a plain term such that $\typjctrl \tpenv \tpenvc \tm
  \tpa$, $\tpenv = \tpenvext {\oftp {\varx_1}{\tpb_1},\:
    \ldots}{\varx_n}{\tpb_n}$, and $\tpenvc = \tpenvext {\oftp
    {\vark_1}{\tpctx {\tpc_1}},\: \ldots}{\vark_m}{\tpctx
    {\tpc_m}}$. Let $\{ \basetpi 1, \ldots, \basetpi p \}=\ftv \tpa
  \cup \ftv \tpenv \cup \ftv \tpenvc$. Let $\vectm \tpd$ be a sequence  of types of length $p$, 
  and let $\vectm \rc$ be reducibility
  candidates such that $\rc_i \in \redc{\tpd_i}$ for all
  $i=1,\ldots,p$. Let $\vectm \tm$ be closed terms such that $\typj
  {\mttpenv;\mttpenvc} {\tm_i}{\subst {\tpb_i}{\vectm \basetp}{\vectm
      \tpd}}$ and $\Qterm {\paramrc{\tpb_i}{\vectm \basetp}{\vectm
      \rc}}{\tm_i}$ for all $i=1,\ldots,n$. Next, let $\vectm {\ctx}$
  be closed contexts such that
  \typj{\mttpenv;\mttpenvc}{\ctx_i}{\tpctx{\tpc_i}} and $\rccont
       {\paramrc {\tpc_i}{\vectm \basetp}{\vectm \rc}}{\ctx_i}$ for
       all $i=1,\ldots,m$, and let $\ctx$ be such that
       \typj{\mttpenv;\mttpenvc}{\ctx}{\tpctx{\tpa}} and $\rccont
            {\paramrc \tpa {\vectm \basetp}{\vectm \rc}} \ctx$. Then
            $\rnorm {\program {\substsimsim \tm {\vectm
                  \basetp}{\vectm \tpd}{\vectm \varx}{\vectm
                  \tm}{\vectm \vark}{\vectm {\ctx}}} \ctx}$ holds.
\end{lemma}

The termination of call-by-name evaluation for well-typed programs
follows from Lemma~\ref{lem:main-cbn-callcc}. As before, the
computational content of the proof takes the form of an evaluator,
only this time in the call-by-name continuation-passing style.

In \cite{Parigot:JSL97}, Parigot gives a proof of strong normalization
for the second-order (\ie\ with the types of System F) $\lambda
\mu$-calculus using a variant of Girard's method of reducibility
candidates. Parigot's proof and ours share some similarities, even
though the results are quite different in nature (strong normalization
vs.  termination of a particular strategy) and Parigot's proof is more
general in that it can be applied to the implicitly typed as well as to the
explicitly typed language. The key point in Parigot's proof is the
following reducibility candidates characterization result: for all
reducibility candidates $\rc$, there exists a set $\mathcal S$ of
(possibly empty) finite sequences of strongly normalizing terms such
that we have $\tm \in \rc$ iff for all $\vectm \tms \in {\mathcal S}$,
$\app \tm {\vectm \tms}$ is strongly normalizing. The greatest such
set $\mathcal S$ is denoted by $\rc^\bot$. The characterization result
is then used to prove a lemma similar to Lemma
\ref{l:termination-cbv-abort}.

The finite sequences of terms can be seen as (call-by-name) contexts
in our setting. Moreover, we notice that $\vectm \tms \in \rc^\bot$
iff for all $\tm \in \rc$, $\app \tm {\vectm \tms}$ is strongly
normalizing; this resembles the definition of the reducibility
predicates on contexts $\rawrccont \rc$ in our proof.  However, the
terms in a sequence $\vectm \tms$ have to be strongly normalizing in
Parigot's proof, while we do not have a similar requirement on
contexts. This fact will have consequences for program extraction; the
program extracted from our proof would be an evaluator in CPS style
where contexts (continuations) are passed around without being
deconstructed (as in~\cite{Biernacka-Biernacki:MFPS09}). 

\section{System F with delimited control operators}
\label{sec:delimited}

In this section, we prove termination of the call-by-value evaluation
in an extension of the explicitly typed System F with the delimited
control operators {\it shift} and {\it reset} of Danvy and
Filinski~\cite{Danvy-Filinski:LFP90}. While the abortive control
operators such as {\it callcc} model jumps, {\it shift} and {\it
  reset} allow for delimited-control capture and continuation
composition.

\subsection{Syntax and semantics}

We extend the explicitly typed System F with the operators {\it shift}
$\rawshift$, {\it reset} $\reset \cdot$, and {\it throw}
$\throw{}{}$. We call the language $\lambdafdel$. The syntax of terms,
term types, contexts, and metacontexts of $\lambdafdel$ is given as
follows:
$$
\hspace{-2mm}
\begin{array}{rrcl}
  \textrm{Terms:} &\tm & ::= & \varx \Mid \lamtp \varx \tpa \tm \Mid \app \tm \tm \Mid \Lam \basetp \tm
  \Mid \apptp \tm \tpa \Mid 
\shift \vark \tm \Mid \reset \tm \Mid \throw \vark
  \tm \Mid \throwctx {\reify \ctx} \tm \\[1mm]
  \textrm{Term types:} &\tpa &::=& \basetp \Mid \arrowtpDF \tpa \tpa \tpa \tpa \Mid \foralltpDF \basetp
  \tpa \tpa \tpa \\[1mm]
  \textrm{CBV contexts:}& \ctx & ::= & \mtctx \Mid \vctx {\lamtpp \varx \tpa \tm}
  \ctx \Mid \apctx \ctx \tm \Mid \aptpctx \ctx \tpa \Mid \throwctx {\reify \ctx} \ctx \\[1mm]
  \textrm{Metacontexts:} & \metactx & ::= & \mtmetactx \Mid \ctxmetactx \ctx \metactx
\end{array}
$$ 
The new term constructs are the {\it shift} operator $\shift\vark\tm$
binding the continuation variable $\vark$ in $\tm$, and a term
delimited by {\it reset}, denoted $\reset\tm$. The remaining term
constructs are as before.  The (non-standard) syntax of types is
discussed in Section \ref{subs:type-system-delimited}. The syntax of
reduction contexts is the same as in Section
\ref{subsec:abortive-syntax-cbv}, and terms are plugged in contexts
using a function $\rawplug$, defined in a similar manner as before.

The new syntactic category is that of metacontexts. A metacontext can
be understood as a stack of contexts: $\mtmetactx$ is the empty
metacontext and the metacontext $\ctxmetactx \ctx \metactx$ is
obtained by pushing the context $\ctx$ on top of $\metactx$ with each
context in the stack separated from the rest by a delimiter. The
meaning of metacontexts is formalized through a function
$\rawplugmeta$, defined below:
\begin{eqnarray*}
  \plugmeta \tm \mtmetactx & = & \tm \\
  \plugmeta \tm {\ctxmetactx \ctx \metactx} &=& \plugmeta {\reset{\plug \tm
      \ctx}} \metactx
\end{eqnarray*}
The result of $\plugmeta \tm \metactx$ is denoted by $\inctx \metactx
\tm$.  

Programs are closed plain terms delimited by a {\it reset}. A program
$\pgm$ is subject to decompositions into a term $t$, a context $\ctx$
and a metacontext $\metactx$ such that $\pgm =
\programDF{\tm}{\ctx}{\metactx}$.

The call-by-value reduction relation on programs in $\lambdafdel$ is defined by
the following rules:
\[
\begin{array}{rcll}
  \programDF {\app {\lamtpp \varx \tpa \tm} \val} \ctx \metactx & \redcbv &  \programDF
  {\subst \tm \varx \val} \ctx \metactx & (\beta_v)\\[2mm]
  \programDF {\apptp {\Lamp \basetp \tm} \tpa} \ctx \metactx & \redcbv &
  \programDF {\subst \tm \basetp \tpa} \ctx \metactx & (\beta_T)\\[2mm]
  \programDF {\shift \vark \tm} \ctx \metactx & \redcbv & 
\program {\reset{\subst \tm
    \vark {\reify \ctx}}} \metactx & (\textit{shift})\\[2mm]
  \programDF {\throw {\reify {\ctx'}} \val} \ctx \metactx & \redcbv & \programDF \val
  {\ctx'} {\ctxmetactx {\ctx}{\metactx}} & (\textit{throw}_v)\\[2mm]
  \programDF {\reset \val} \ctx \metactx & \redcbv & \programDF \val \ctx
  \metactx & (\textit{reset})
\end{array}
\]
where values are defined as before. The first two reduction rules are
standard (and insensitive to the surrounding context and metacontext).
The rule $(\textit{shift})$ states that reducing $\shift \vark \tm$
consists in capturing the context $\ctx$ and substituting it for the
continuation variable $\vark$ in the body $\tm$ (the current context
is then set to be empty). When a captured context $\ctx'$ is applied
to a value (in the rule $(\textit{throw}_v)$), it is reinstated as the
current context, and the then-current context $\ctx$ is pushed on the
metacontext $\metactx$. Finally, the last rule states that when a
value is enclosed in a {\it reset}, it means that the {\it reset} can
be discarded since no further captures can occur inside it.

The evaluation relation $\redcbv^*$ is defined as before, where the
expected result of evaluation is a program value of the form
$\reset{v}$.

\subsection{Type system}
\label{subs:type-system-delimited}

\begin{figure}[t!!]
\begin{flushleft}
Typing terms ($\tpa ::= \basetp \Mid \arrowtpDF \tpa \tpb \tpc \tpd \Mid \foralltpDF \basetp \tpa \tpb \tpc$):
\end{flushleft} 
\begin{mathpar}
  \inferrule{ }
  {\typejDF{\tpenvext \tpenv \varx \tpa} \tpenvc \tpb \varx \tpa \tpb}
  \and
  \inferrule{\typejDF{\tpenvext \tpenv \varx \tpa} \tpenvc \tpc \tm \tpb \tpd}
  {\typejDF \tpenv \tpenvc \tpe {\lamtp \varx \tpa \tm}{\arrowtpDF \tpa \tpb
      \tpc \tpd} \tpe}
  \and
  \inferrule{\typejDF \tpenv \tpenvc \tpf \tmzero {\arrowtpDF
      \tpa \tpb \tpc \tpe} \tpd \\ \typejDF \tpenv \tpenvc \tpe \tmone \tpa \tpf}
  {\typejDF \tpenv \tpenvc \tpc {\app \tmzero \tmone}{\tpb} \tpd}
  \and
  \inferrule{\typejDF \tpenv \tpenvc \tpb \tm \tpa \tpc \\ \basetp \notin
    \ftv \tpenv \cup \ftv \tpenvc}
  {\typejDF \tpenv \tpenvc \tpd {\Lam \basetp \tm} {\foralltpDF \basetp \tpa \tpb \tpc} \tpd}
  \and
  \inferrule{\typejDF \tpenv \tpenvc {\subst \tpc \basetp \tpd} \tm {\foralltpDF
      \basetp \tpa \tpb \tpc} \tpe}
  {\typejDF \tpenv \tpenvc {\subst \tpb \basetp \tpd}{\apptp \tm \tpd}{\subst
      \tpa \basetp \tpd} \tpe}
  \and
  \inferrule{\typejDF \tpenv \tpenvc \tpc \tm \tpc \tpa}
  {\typejDF \tpenv \tpenvc \tpb {\reset \tm} \tpa \tpb}
  \and
  \inferrule{\typejDF \tpenv {\tpenvext \tpenvc \vark {\tpctxDF \tpa \tpb}} \tpd \tm \tpd \tpc}
  {\typejDF \tpenv \tpenvc \tpb {\shift \vark \tm} \tpa \tpc}
  \and
  \hspace{-5mm}\inferrule{\typejDF \tpenv {\tpenvext \tpenvc \vark {\tpctxDF \tpa \tpb}} \tpc \tm \tpa \tpd}
  {\typejDF \tpenv {\tpenvext \tpenvc \vark {\tpctxDF \tpa \tpb}} \tpc
    {\throw \vark \tm} \tpb \tpd}
  \and
  \inferrule{\typejcDF \tpenv \tpenvc \ctx {\tpctxDF \tpa \tpb} \\
    \typejDF \tpenv \tpenvc \tpc \tm \tpa \tpd}
  {\typejDF \tpenv \tpenvc \tpc {\throw {\reify \ctx} \tm} \tpb \tpd}
\end{mathpar}
\begin{flushleft}
Typing contexts ($\tpcont ::= \tpctxDF \tpa \tpb$): 
\end{flushleft}
\begin{mathpar}
  \inferrule{ }
  {\typejcDF \tpenv \tpenvc \mtctx {\tpctxDF \tpa \tpa}}
  \and
  \inferrule{\typejcDF \tpenv \tpenvc \ctx {\tpctxDF \tpb \tpc} \\
    \typejDF \tpenv \tpenvc \tpd \tm \tpa \tpe}
  {\typejcDF \tpenv \tpenvc {\apctx \ctx \tm}{\tpctxDF {(\arrowtpDF \tpa \tpb
        \tpc \tpd)} \tpe}}
  \and
  \inferrule{\typejDF \tpenv \tpenvc \tpe {\lamtp \varx \tpa \tm}{\arrowtpDF \tpa \tpb
        \tpc \tpd} \tpe \\
   \typejcDF \tpenv \tpenvc \ctx {\tpctxDF \tpb \tpc}}
  {\typejcDF \tpenv \tpenvc {\vctx {\lamtpp \varx \tpa \tm} \ctx}{\tpctxDF
      \tpa \tpd}}
  \and
  \inferrule{\typejcDF \tpenv \tpenvc \ctx {\tpctxDF {\subst \tpa \basetp
        \tpd}{\subst \tpb \basetp \tpd}}}
  {\typejcDF \tpenv \tpenvc {\aptpctx \ctx \tpd}{\tpctxDF {\foralltpDF
        \basetp \tpa \tpb \tpc}{\subst \tpc \basetp \tpd}}}  
  \and
  \inferrule{\typejcDF \tpenv \tpenvc {\ctx'}{\tpctxDF \tpa \tpb} \\
    \typejcDF \tpenv \tpenvc \ctx {\tpctxDF \tpb \tpc}}
  {\typejcDF \tpenv \tpenvc {\throw {\reify {\ctx'}} \ctx}{\tpctxDF \tpa \tpc}}
\end{mathpar}
\begin{flushleft}
Typing metacontexts ($\tpmcont ::= \tpmetactxDF \tpa$): 
\end{flushleft}
\begin{mathpar}
  \inferrule{ }
  {\typejcDF \tpenv \tpenvc \mtmetactx {\tpmetactxDF \tpa}}
  \and
  \inferrule{\typejcDF \tpenv \tpenvc \ctx {\tpctxDF \tpa \tpb} \\
    \typejcDF \tpenv \tpenvc \metactx {\tpmetactxDF \tpb}}
  {\typejcDF \tpenv \tpenvc {\ctxmetactx \ctx \metactx}{\tpmetactxDF \tpa}}
\end{mathpar}
\caption{Typing rules for System F with delimited control operators under call by value}
\label{f:type-system-delimited}
\end{figure}

We add System F types to the type system of Biernacka and
Biernacki~\cite{Biernacka-Biernacki:PPDP09}, which is a slight
modification of the classical Danvy and Filinski's type system for
{\it shift} and {\it reset}~\cite{Danvy-Filinski:DIKU89}. The type
system is presented in Figure~\ref{f:type-system-delimited}. In a type
$\foralltpDF \basetp \tpa \tpb \tpc$, the quantifier binds the
occurrences of $\basetp$ in $\tpa, \tpb$, and $\tpc$. We define the
set of free type variables $\ftv \tpa$ of a term type $\tpa$
accordingly, and we define $\ftv {\tpctxDF \tpa \tpb} \isdef \ftv \tpa
\cup \ftv \tpb$.

In this system, contexts are assigned types of the form $\tpctxDF \tpa
\tpb$, where $\tpa$ is the type of the hole and $\tpb$ is the answer
type, and metacontexts are assigned types of the form $\tpmetactxDF
\tpa$, where $\tpa$ is the type of the hole. A typing judgment
$\typejDF \tpenv \tpenvc \tpb \tm \tpa \tpc$ roughly means that under
the assumptions $\tpenv$ and $\tpenvc$, the term $\tm$ can be plugged
into a context of type $\tpctxDF \tpa \tpb$ and a metacontext of type
$\tpmetactxDF \tpc$ (in general, the evaluation of $\tm$ may use the
surrounding context of type $\tpctxDF \tpa \tpb$ to produce a value of
type $\tpc$, with $\tpb \neq \tpc$). Because both abstractions $\lamtp
\varx \tpa \tm$ and $\Lam \basetp \tm$ denote ``frozen''
computations---waiting for a term and a type, resp., to activate
them---the arrow type and the $\forall$-type contain additional type
annotations. Roughly, the type $\arrowtpDF \tpa \tpb \tpc \tpd$ is
assigned to a function that can be applied to an argument of type
$\tpa$ within a context of type $\tpctxDF \tpb \tpc$ and a metacontext
of type $\tpmetactxDF \tpd$. Similarly, the type $\foralltpDF \basetp
\tpa \tpb \tpc$ is assigned to a term that can be applied to a type
$\tpd$ within a context of type $\tpctxDF{\subst \tpa \basetp
  \tpd}{\subst \tpb \basetp \tpd}$ and a metacontext of type
$\tpmetactxDF {\subst \tpc \basetp \tpd}$. It can be shown that closed
well-typed terms either are values or decompose uniquely into a redex,
a context and a metacontext, and that the reduction rules preserve
types.

The type system of Figure~\ref{f:type-system-delimited} is more
liberal than the one defined for $\lambda_2^{s/r,Std}$, a language
defined by Asai and Kameyama in~\cite{Asai-Kameyama:APLAS07}, which is
similar to $\lambdafdel$. In~\cite{Asai-Kameyama:APLAS07}, polymorphic
abstraction types do not contain any additional type annotations, and
can only be assigned to abstractions $\Lam \basetp \tm$ where $\tm$ is
a \emph{pure} term, \ie\ a term such that $\typejDF \tpenv \tpenvc
\tpb \tm \tpa \tpb$ is derivable for any $\tpb$. Pure terms are terms
free from control effects, such as $\varx$, $\lamtp \varx \tpa \tm$,
$\reset \tm$, or $\Lam \basetp \tm$. In contrast, we allow arbitrary
abstractions of the form $\Lam \basetp \tm$, at the cost of additional
type annotations in the polymorphic abstraction types. As pointed out
by Asai and Kameyama, restricting $\forall$-introduction to pure terms
is not mandatory in a calculus with standard call-by-value evaluation,
such as $\lambda_2^{s/r,Std}$ and our calculus. However, such
restriction becomes necessary for the calculus $\lambda_2^{s/r,ML}$ of
\cite{Asai-Kameyama:APLAS07} with ML-like call-by-value evaluation
(where reduction is allowed under $\Lambda$) for subject reduction to
hold.

\subsection{Termination}

The proof of termination is very similar to that of
Section~\ref{subsec:termination-abort}, and here we only point out the
main differences. This time our development is based on the layered
continuation-passing style for {\it shift} and {\it reset}, where
terms are passed two layers of
continuations~\cite{Danvy-Filinski:LFP90}.

We define the normalization predicate $\rnorm \pgm$
as follows:
$$\rnorm \pgm \isdef \Exists \val {\pgm \relevalcbv \reset{\val}}$$ 
A \emph{reducibility candidate} $\rc$ of type
$\tpa$ is a set of closed values of type $\tpa$. We write $\redc \tpa$
for the set of reducibility candidates of type $\tpa$. Let $\rc$,
$\rcs$ be reducibility candidates of types $\tpa$ and $\tpb$,
respectively. We define the predicate $\rccontDF \rc \rcs \ctx$ on
closed contexts of type $\tpctxDF \tpa \tpb$ and the predicate
$\rcmetacont \rc \metactx$ on closed metacontexts of type
$\tpmetactxDF \tpa$ as follows:
\begin{eqnarray*}
  \rccontDF \rc \rcs \ctx &\isdef& \Forall \val {\val \in \rc \impl {\Forall
      \metactx {\rcmetacont \rcs \metactx \impl \rnorm {\programDF \val
          \ctx \metactx}}}}\\
  \rcmetacont \rc \metactx &\isdef& \Forall \val {\val \in \rc \impl \rnorm 
{\inctx{\metactx}{\reset{\val}}}}
\end{eqnarray*}
Let $\tpa$ be a type, $\ftv \tpa \subseteq \vectm \basetp$, $\vectm \tpb$ be a
sequence of types of the same size as $\vectm \basetp$, and $\vectm
\rc$ be reducibility candidates such that each $\rc_i$ is of type
$\tpb_i$. We now define the parametric reducibility candidate $\paramrc
\tpa {\vectm \basetp}{\vectm \rc}$ of type $\subst \tpa {\vectm
  \basetp}{\vectm \tpb}$ as follows:
\[
\begin{array}{rcl}
  \val \in \paramrc {\basetp_i}{\vectm \basetp}{\vectm \rc} 
  & \mbox{ iff } 
  & \val \in \rc_i 
  \\
  \valzero \in \paramrc {\arrowtpDF \tpa \tpb \tpc \tpd}{\vectm \basetp}{\vectm \rc}
  & \mbox{ iff } 
  & \forall \valone . \valone \in \paramrc {\tpa}{\vectm \basetp}{\vectm \rc}
    \impl 
    \forall \ctx. \rccontDF {\paramrc \tpb {\vectm \basetp}{\vectm
          \rc}}{\paramrc \tpc {\vectm \basetp}{\vectm \rc}} \ctx 
    \impl 
\\ && \hspace{4mm} 
    \forall \metactx . \rcmetacont {\paramrc \tpd {\vectm \basetp}{\vectm
            \rc}} \metactx \impl
    \rnorm {\programDF {\app \valzero \valone} \ctx
        \metactx}
  \\
  \val \in \paramrc {\foralltpDF \basetpc \tpa \tpb \tpc}{\vectm \basetp}{\vectm \rc} &\mbox
  { iff }& \forall
  \tpd. \forall \rcs. \rcs \in \redc \tpd \impl 
  \forall \ctx. \rccontDF
        {\paramrcsim \tpa {\vectm \basetp}{\vectm \rc} \basetpc \rcs}
        {\paramrcsim \tpb {\vectm \basetp}{\vectm \rc} \basetpc \rcs} \ctx \impl 
\\ && \hspace{4mm}
        \forall \metactx. \rcmetacont {\paramrcsim \tpc {\vectm \basetp}{\vectm
              \rc} \basetpc \rcs} \metactx \impl 
        \rnorm {\programDF {\apptp \val
              \tpd} \ctx \metactx}
\end{array}
\]
Using a substitution lemma similar to Lemma \ref{l:paramrc-subst}, we
can prove the following result, from which the termination theorem
follows for closed plain terms:
\begin{lemma}
\label{lem:termination-cbv-shift}
  Let $\tm$ be a plain term such that $\typejDF {\tpenv}{\tpenvc} \tpb
  \tm \tpa \tpc$, $\tpenv = \tpenvext {\oftp {\varx_1}{\tpb_1},\:
    \ldots}{\varx_n}{\tpb_n}$, $\tpenvc = \tpenvext {\oftp
    {\vark_1}{\tpcont_1},\: \ldots}{\vark_m}{\tpcont_m}$. Let $\{
  \basetp_1, \ldots, \basetp_p \} = \ftv \tpa \cup \ftv{\tpb} \cup
  \ftv{\tpc} \cup \ftv \tpenv \cup \ftv\tpenvc$. Let $\vectm \tpd$ be
  a sequence of types of length $p$, and let $\vectm \rc$ be
  reducibility candidates such that $\rc_i \in \redc{\tpd_i}$. Let
  $\vectm \val$ be closed values such that $\typejDF
  {\mttpenv}{\mttpenvc} \tpe {\val_i}{\subst {\tpb_i}{\vectm
      \basetp}{\vectm \tpd}} \tpe$ and $\val_i \in
  \paramrc{\tpb_i}{\vectm \basetp}{\vectm \rc}$ for each $i$. Let
  $\vectm {\ctx}$ be closed contexts such that $\typejcDF \mttpenv
  \mttpenvc {\ctx_i}{\tpcont_i}$, $\tpcont_i = \tpctxDF
            {\tpe_i^1}{\tpe_i^2}$, and $\rccontDF {\paramrc
              {\tpe_i^1}{\vectm \basetp}{\vectm \rc}}{\paramrc
              {\tpe_i^2}{\vectm \basetp}{\vectm \rc}}{\ctx_i}$ for
            each $i$. Let $\ctx$ be such that $\typejcDF \mttpenv
            \mttpenvc {\ctx}{\tpctxDF \tpa \tpb}$ and $\rccontDF
                      {\paramrc \tpa {\vectm \basetp}{\vectm
                          \rc}}{\paramrc \tpb {\vectm \basetp}{\vectm
                          \rc}} \ctx$. Let $\metactx$ be such that
                      $\typejcDF \mttpenv \mttpenvc
                      {\metactx}{\tpmetactxDF \tpc}$ and $\rcmetacont
                      {\paramrc \tpc {\vectm \basetp}{\vectm \rc}}
                      \metactx$. Then $\rnorm {\programDF
                        {\substsimsim \tm {\vectm \basetp}{\vectm
                            \tpd}{\vectm \varx}{\vectm \val}{\vectm
                            \vark}{\vectm {\ctx}}} \ctx \metactx}$
                      holds.
\end{lemma}

\begin{theorem}
\label{thm:termination-cbv-shift}
  If $\pgm$ is a well-typed program, then $\rnorm{\pgm}$ holds.
\end{theorem}

\proof We have $\rccontDF{\rc}{\rc}{\mtctx}$ for any $\rc$ and
$\rcmetacont{\rc}{\mtmetactx}$ for any $\rc$, so we can use the
previous lemma.  \qed

Again, the computational content of this proof takes the form of an
evaluator in CPS, this time with two layers of continuations
~\cite{Biernacka-Biernacki:PPDP09}.

\subsection{Call by name}

The developments of the previous section can be easily adapted to the
call-by-name strategy. Following~\cite{Biernacka-Biernacki:PPDP09}, we
modify the syntax of types to express the fact that functions accept
not values, but suspended computations expecting a continuation:
\[
\begin{array}{rrcl}
  \textrm{Terms:} &\tm & ::= & \varx \Mid \lamtp \varx {\tpvarcbn \tpa \tpa \tpa} \tm \Mid \ldots \\[1mm]
  \textrm{Term types:} &\tpa &::=& \basetp \Mid \arrowtpDF {\tpvarcbn \tpa \tpa \tpa} \tpa \tpa \tpa \Mid \foralltpDF \basetp
  \tpa \tpa \tpa 
\end{array}
\]

The syntax of reduction contexts is modified:
\[
\begin{array}{rrcl}
  \textrm{CBN contexts:}& \ctx & ::= & \mtctx \Mid \apctx \ctx \tm \Mid \aptpctx \ctx \tpa 
\end{array}
\]

The reduction rules are modified as in Section~\ref{subsec:callbyname-abort}:
\[
\begin{array}{rcll}
  \programDF {\app {\lamtpp \varx {\tpvarcbn \tpa \tpb \tpc} \tmzero} \tmone} \ctx \metactx & \redcbn &  \programDF
  {\subst \tmzero \varx \tmone} \ctx \metactx & (\beta_n)\\[2mm]
  \programDF {\apptp {\Lamp \basetp \tm} \tpa} \ctx \metactx & \redcbn &
  \programDF {\subst \tm \basetp \tpa} \ctx \metactx & (\beta_T)\\[2mm]
  \programDF {\shift \vark \tm} \ctx \metactx & \redcbn &   
\program {\reset{\subst \tm
    \vark {\reify \ctx}}} \metactx
& (\textit{shift})\\[2mm]
  \programDF {\throw {\reify {\ctx'}} \tm} \ctx \metactx & \redcbn & \programDF \tm
  {\ctx'} {\ctxmetactx {\ctx}{\metactx}} & (\textit{throw}_n)\\[2mm]
  \programDF {\reset \val} \ctx \metactx & \redcbn & \programDF \val \ctx
  \metactx & (\textit{reset})
\end{array}
\]

The typing rules for abstractions, function applications and throwing
are changed, as in~\cite{Biernacka-Biernacki:PPDP09}. The modified
rules for abstractions and function applications are as follows:
\begin{mathpar}
  \inferrule{\typejDF{\tpenvext \tpenv \varx {\tpvarcbn \tpa \tpb \tpc}} \tpenvc \tpe \tm \tpd \tpf}
  {\typejDF \tpenv \tpenvc \tpg {\lamtp \varx {\tpvarcbn \tpa \tpb \tpc} \tm}{\arrowtpDF {\tpvarcbn \tpa \tpb \tpc} \tpd
      \tpe \tpf} \tpg}
  \and
  \inferrule{\typejDF \tpenv \tpenvc \tpf \tmzero {\arrowtpDF
      {\tpvarcbn \tpa \tpb \tpc} \tpd \tpe \tpf} \tpg \\ \typejDF \tpenv \tpenvc \tpb \tmone \tpa \tpc}
  {\typejDF \tpenv \tpenvc \tpe {\app \tmzero \tmone}{\tpd} \tpg}
\end{mathpar}
We also modify the rules for typing contexts in a straightforward way.

We define reducibility candidates and the predicates $\rawrccontDF \rc
\rcs$ and $\rawrcmetacont \rct$ as in call by value, the predicate
$\rnorm \cdot$ is defined using $\redcbn$, the call by name reduction
relation. The case for function types in the definition of the
parametric reducibility candidates is as in
Section~\ref{subsec:callbyname-abort}. The definition of the
$\rawQtermDF \rc \rcs \rct$ is adapted to the language with {\it
  shift} and {\it reset}.
\[
\begin{array}{rcl}
  \val \in \paramrc {\basetp_i}{\vectm \basetp}{\vectm \rc} 
  & \mbox{ iff } 
  & \val \in \rc_i 
  \\
  \val \in \paramrc {\arrowtpDF {\tpvarcbn \tpa \tpb \tpc} \tpd \tpe \tpf}{\vectm \basetp}{\vectm \rc}
  & \mbox{ iff } 
  & \forall \tm . \QtermDF {\paramrc {\tpa}{\vectm \basetp}{\vectm \rc}} 
        {\paramrc {\tpb}{\vectm \basetp}{\vectm \rc}} 
        {\paramrc {\tpc}{\vectm \basetp}{\vectm \rc}} {\tm} 
        \impl 
        \QtermDF {\paramrc {\tpd}{\vectm \basetp}{\vectm \rc}}
            {\paramrc {\tpe}{\vectm \basetp}{\vectm \rc}}
            {\paramrc {\tpf}{\vectm \basetp}{\vectm \rc}} {\app \val \tm}
  \\
  \val \in \paramrc {\foralltpDF \basetpc \tpa \tpb \tpc}{\vectm \basetp}{\vectm \rc} &\mbox
  { iff }& \forall
  \tpd. \forall \rcs. \rcs \in \redc \tpd \impl 
  \forall \ctx. \rccontDF
        {\paramrcsim \tpa {\vectm \basetp}{\vectm \rc} \basetpc \rcs}
        {\paramrcsim \tpb {\vectm \basetp}{\vectm \rc} \basetpc \rcs} \ctx \impl 
\\ && \hspace{4mm}
        \forall \metactx. \rcmetacont {\paramrcsim \tpc {\vectm \basetp}{\vectm
              \rc} \basetpc \rcs} \metactx \impl 
        \rnorm {\programDF {\apptp \val
              \tpd} \ctx \metactx}
\end{array}
\]
with
\[
\QtermDF \rc \rcs \rct \tm  = \Forall \ctx {\rccontDF \rc \rcs \ctx \impl 
    \Forall \metactx {\rcmetacont \rct \metactx \impl
    \rnorm {\programDF \tm \ctx \metactx}}}
\]

The main lemma is as follows:
\begin{lemma}
  Let $\tm$ be a plain term such that $\typejDF {\tpenv}{\tpenvc} \tpb
  \tm \tpa \tpc$, $\tpenv = \tpenvext {\oftp {\varx_1}{\tpvarcbn
      {\tpa_1} {\tpb_1} {\tpc_1}},\: \ldots}{\varx_n}{\tpvarcbn
    {\tpa_n} {\tpb_n} {\tpc_n}}$, $\tpenvc = \tpenvext {\oftp
    {\vark_1}{\tpcont_1},\: \ldots}{\vark_m}{\tpcont_m}$. Let $\{
  \basetp_1, \ldots, \basetp_p \} = \ftv \tpa \cup \ftv{\tpb} \cup
  \ftv{\tpc} \cup \ftv \tpenv \cup \ftv\tpenvc$. Let $\vectm \tpd$ be
  a sequence of types of length $p$, and let $\vectm \rc$ be
  reducibility candidates such that $\rc_i \in \redc{\tpd_i}$. Let
  $\vectm \tmr$ be closed terms such that $\typejDF
  {\mttpenv}{\mttpenvc} {\subst {\tpb_i} {\vectm \basetp} {\vectm
      \tpd}} {\tmr_i}{\subst {\tpa_i}{\vectm \basetp}{\vectm \tpd}}
  {\subst {\tpc_i} {\vectm \basetp} {\vectm \tpd}}$ and $\QtermDF
  {\paramrc{\tpa_i}{\vectm \basetp}{\vectm \rc}}
  {\paramrc{\tpb_i}{\vectm \basetp}{\vectm \rc}}
  {\paramrc{\tpc_i}{\vectm \basetp}{\vectm \rc}} {\tmr_i}$ for each
  $i$. Let $\vectm {\ctx}$ be closed contexts such that $\typejcDF
  \mttpenv \mttpenvc {\ctx_i}{\tpcont_i}$, $\tpcont_i = \tpctxDF
           {\tpe_i^1}{\tpe_i^2}$, and $\rccontDF {\paramrc
             {\tpe_i^1}{\vectm \basetp}{\vectm \rc}}{\paramrc
             {\tpe_i^2}{\vectm \basetp}{\vectm \rc}}{\ctx_i}$ for each
           $i$. Let $\ctx$ be such that $\typejcDF \mttpenv \mttpenvc
           {\ctx}{\tpctxDF \tpa \tpb}$ and $\rccontDF {\paramrc \tpa
             {\vectm \basetp}{\vectm \rc}}{\paramrc \tpb {\vectm
               \basetp}{\vectm \rc}} \ctx$. Let $\metactx$ be such
           that $\typejcDF \mttpenv \mttpenvc {\metactx}{\tpmetactxDF
             \tpc}$ and $\rcmetacont {\paramrc \tpc {\vectm
               \basetp}{\vectm \rc}} \metactx$. Then $\rnorm
           {\programDF {\substsimsim \tm {\vectm \basetp}{\vectm
                 \tpd}{\vectm \varx}{\vectm \tmr}{\vectm \vark}{\vectm
                 {\ctx}}} \ctx \metactx}$ holds.
\end{lemma}

\section{Conclusion and perspectives}
\label{sec:conclusion}

We have shown that the context-based proof method developed by the
first two authors for the simply-typed lambda calculus with control
operators, be they abortive or delimited, scales to much more
expressive type systems based on System F. The presented proofs are
rather simple and elegant, and they do not require a journey through
an optimized CPS translation in order to show termination of
evaluation for such calculi. Furthermore, if formalized in a logical
framework equipped with program extraction mechanism, they can lead to
executable specifications of programming languages with control
operators and polymorphism---which is left as future work.

The proof method we have proposed is tailored towards characterization
of termination in a wide range of context-sensitive reduction
semantics that account for arbitrary reduction strategies or advanced
control operators. For example, the proof of termination for
delimited-control operators of Section~\ref{sec:delimited} can be
straightforwardly generalized to a polymorphic version of the CPS
hierarchy of Danvy and
Filinski~\cite{Biernacka-al:PPDP11,Danvy-Filinski:LFP90}. Similarly,
we expect that one could consider a call-by-need version of System F,
e.g., based on~\cite{Danvy-al:FLOPS10} and readily apply the
context-based method to it. Such results do not seem to be immediately
obtainable in other frameworks, \eg, using orthogonality techniques.

{\bf Acknowledgments:} We thank the anonymous reviewers for detailed
and insightful comments on several versions of this article. This work
has been partially supported by Polish NCN grant number
DEC-011/03/B/ST6/00348.

\bibliographystyle{abbrv}

\end{document}